\documentclass[aps, 11pt, preprint, prc, nofootinbib, superscriptaddress]{revtex4}

\usepackage[usenames]{color}
\usepackage{graphicx}
\usepackage{amsmath}
\usepackage{amsfonts}
\usepackage{amssymb}
\usepackage{mathrsfs}
\usepackage{bm}
\usepackage{verbatim}
\usepackage{cancel}
\usepackage{CJK}

\usepackage{comment}
\usepackage[normalem]{ulem}
\usepackage[dvipsnames]{xcolor}
\usepackage[utf8]{inputenc}

\newcommand{\mhi}{M_{\text{hi}}}

\newcommand{\chn}[3]{{{}^{#1}\!{#2}_{#3}}}
\newcommand{\cs}[2]{\chn{#1}{S}{#2}}
\newcommand{\cp}[2]{\chn{#1}{P}{#2}}
\newcommand{\cd}[2]{\chn{#1}{D}{#2}}
\newcommand{\cf}[2]{\chn{#1}{F}{#2}}

\newcommand{\csd}{{\cs{3}{1}-\cd{3}{1}}}
\newcommand{\cpf}{{\cp{3}{2}-\cf{3}{2}}}

\newcommand{\NNLO}{N$^2$LO}
\newcommand{\NNNLO}{N$^3$LO}

\begin{document}

\title{Perturbative chiral nucleon-nucleon potential for the $^3P_0$ partial wave}

\author{Rui Peng}

\author{Songlin Lyu}

\author{Bingwei Long}
\email{bingwei@scu.edu.cn}
\affiliation{College of Physics, Sichuan University, Chengdu, Sichuan 610065, China}

\date{May 23, 2020}

\begin{abstract}
We study perturbativeness of chiral nuclear forces in the $^3P_0$ channel. In previous works, the focus has been on one-pion exchange, and the applicable window of perturbative pion exchanges has been shown to span from threshold to the center-of-mass momentum $k \simeq$ 180 MeV. We will examine, instead, whether cancellation of short and long-range parts can sufficiently soften the $^3P_0$ chiral force to make it more amenable to perturbation theory. The result is encouraging, as the combined $^3P_0$ force is shown to be perturbative up to $k \simeq$ 280 MeV, covering many nuclear-structure calculations.

\end{abstract}

\maketitle

\section{Introduction\label{sec:intro}}

In the framework of chiral effective field theory (EFT), one-pion exchange (OPE) is the leading-order (LO) long-range part of two-body nuclear forces. Not only does it naturally give rise to tensor structure of nuclear forces, which is well known to play an important role in accounting for nuclear phenomena, but renormalization of OPE tells a lot about short-range nuclear forces~\cite{Kaplan:1998we,Kaplan:1998tg}. However, whether OPE is perturbative must first be decided and renormalization of OPE can be qualitatively different for perturbative and nonperturbative scenarios~\cite{Long:2007vp,Nogga:2005hy,Valderrama:2009ei,PavonValderrama:2010fb,PavonValderrama:2011zz,Valderrama:2011mv,Long:2011qx,Long:2011xw,Long:2012ve, Ren:2016jna, Wang:2020myr}. It may be desirable for some of the few-body or many-body methods to have perturbation theory as widely applicable as possible, so that computational model space is smaller at least for LO.

Suppression by centrifugal barriers has been the main mechanism to render OPE perturbative; therefore, power counting of chiral nuclear forces has been discussed by partial wave. It has been established that OPE is perturbative in ${}^3P_0$ up to $k \simeq 180$ MeV~\cite{Birse:2005um,Wu:2018lai,Kaplan:2019znu}, where $k$ is the center-of-mass momentum. (For such low momenta, one might as well resort to pionless theory.) On the other hand, the $\cp{3}{0}$ phase shifts are much smaller than those of $\cs{1}{0}$ and $\csd$, rising to the maximum of 10.8$^\circ$ around $k \simeq 160$ MeV and vanishing around $k \simeq 295$ MeV. One may be wondering why no perturbation theory would work for momenta higher than $k \simeq 180$ MeV. In the present paper, we consider a novel perturbative scenario for $^3P_0$: singular attractive OPE and the $^3P_0$ short-range force, repulsive for soft cutoff values, cancel each other partially to bring about a smaller total force.

The singular attraction of OPE in some of the triplet channels, e.g., $\csd$, $
{}^3P_0$, and $\cpf$, etc., does not stabilize $NN$ system by itself, so a
counterterm is needed at LO to make the problem well-defined. Systematic
investigation of how long-range forces affect the role of short-range
counterterms is often recast in the language of renormalization-group (RG)
analysis. Renormalization-group analysis studies the dependence of observables on the ultraviolet cutoff in the dynamic equation. High sensitivity to the cutoff value, if any, indicates the lack of our understanding of short-range physics; therefore, more low-energy constants in the effective Lagrangian must be called upon to parametrize short-range forces. We will not focus on the topic of renormalization in the paper, but the lessons learned so far will be applied to study perturbativeness of the ${}^3P_0$ nuclear force.

By naive dimensional analysis (NDA), the ${}^3P_0$ counterterm is at most next-to-next-to-leading order (\NNLO), suppressed by $\mathcal{O}(Q^2/\mhi^2)$ compared to LO. On the ground of renormalization, however, it is promoted to LO \cite{Nogga:2005hy} (for a more comprehensive review on renormalization of chiral nuclear forces, see ~\cite{vanKolck:2020llt}). For relatively small cutoff values, the ${}^3P_0$ counterterm is repulsive~\cite{Birse:2005um,Nogga:2005hy}, so we are motivated to look into whether the attraction of OPE and the repulsion of the ${}^3P_0$ counterterm can combine to form so small a sum that perturbation theory for ${}^3P_0$ is viable for nuclear physics.

The paper is structured as follows. In Sec.~\ref{sec:pc} we explain the
related power counting schemes, followed by numerical results and comparisons to the empirical phase shifts in Sec.~\ref{sec:results}. Finally, we summarize and discuss in Sec.~\ref{sec:sum}.

\section{Power counting\label{sec:pc}}

This is probably a good place to formulate in power-counting language what we mean by ``perturbation theory''. LO is always nonperturbative, i.e., the LO potential being iterated to all orders using the Lippmann-Schwinger(LS) equation:
\begin{equation}
T^{(0)} =  V_\text{LO} + V_\text{LO}G_0T^\text{(0)} = V_\text{LO} + V_\text{LO}G_0V_\text{LO} + V_\text{LO}(G_0V_\text{LO})^2 + ... \, , \label{LS}
\end{equation}
where $G_0$ is the free-particle propagator. More specifically, the integral equation for uncoupled partial waves to implement such iterations is given by
\begin{equation}
  T^{(0)}(p', p; k) = V^\Lambda_\text{LO}(p', p) + \int_0^{\infty} {\rm d} l\, l^2\, V^\Lambda_\text{LO}(p', l; k) \frac{T^{(0)}(l, p; k)}{k^2 - l^2 + {\rm i}\epsilon} \, ,
\end{equation}
where full off-shell form of partial-wave amplitude $T$ is shown, $p$ ($p'$) is incoming (outgoing) momentum, and $k$ is related to the center-of-mass energy $E$ by the usual nonrelativistic kinematics:
\begin{equation}
  k^2 = 2 m_N E \, ,
\end{equation}
where the nucleon mass $m_N$ = $938.27$ MeV. The amplitude is physical when momenta are on-shell $p' = p = k$: $T(k, k; k)$. The superscript $\Lambda$ is there to remind us that $V^\Lambda_\text{LO}$ has ultraviolet regularization as follows:
\begin{equation}
  V^\Lambda_\text{LO}(p', p) = \exp\left(-\frac{{p'}^4}{\Lambda^4}\right) V_\text{LO}(p', p) \exp\left(-\frac{p^4}{\Lambda^4}\right) \, .
\end{equation}
 Normalization of $T$ and $V$ can be inferred from the integral equation and it is consistent with the following relation between $T(k, k; k)$ and the $\cp{3}{0}$ phase shifts:
\begin{equation}
   \langle \cp{3}{0} | T(k) |\cp{3}{0} \rangle = - \frac{2}{\pi} \frac{{\rm e}^{{\rm i}\delta} \sin \delta}{k} \, ,
\label{eqn:defTmat}
\end{equation}
where $\delta$ is the phase shift at $k$.

NLO and higher orders are perturbations on top of LO. Defining operator function,
\begin{equation}
  F(V) \equiv (1+T^{(0)}G_0)V(1+G_0T^{(0)}) \, ,\label{Fv}
\end{equation}
we can write subleading corrections to the amplitude as
\begin{equation}
  \begin{aligned}
    T^{(1)} &=  F(V_\text{NLO})\, , \\
    T^{(2)} &=  F(V_\text{N$^2$LO}) + F(V_\text{NLO}G_0V_\text{NLO}) + F(V_\text{NLO}G_0T^{(0)}G_0V_\text{NLO})\, , \\
    T^{(3)} &=  F(V_\text{N$^3$LO}) + F(V_\text{N$^2$LO}G_0V_\text{NLO}) + F(V_\text{NLO}G_0V_\text{N$^2$LO}) + F(V_\text{NLO}G_0V_\text{NLO}G_0V_\text{NLO}) \\
    & + F(V_\text{NLO}G_0T^{(0)}G_0V_\text{NLO}G_0V_\text{NLO}) + F(V_\text{NLO}G_0V_\text{NLO}G_0T^{(0)}G_0V_\text{NLO}) \\
    & + F(V_\text{NLO}G_0T^{(0)}G_0V_\text{NLO}G_0T^{(0)}G_0V_\text{NLO}) + F(V_\text{N$^2$LO}G_0T^{(0)}G_0V_\text{NLO})\\ &
    + F(V_\text{NLO}G_0T^{(0)}G_0V_\text{N$^2$LO}) \, , \\
    & \cdots
  \end{aligned}\label{eqn:amplitude}
\end{equation}
What we mean by ``perturbation theory'' is when $V_\text{LO}$ vanishes, therefore, the amplitude consists of only a finite orders of the Born expansion terms.

The OPE potential ~\cite{Bedaque-2002mn, Epelbaum-2008ga, Machleidt-2011zz} is given by
\begin{equation}
  V_{1\pi}(\vec{q}\,) = - \frac{g_A^2}{4f_\pi^2} \bm{\tau}_1 \bm{\cdot} \bm{\tau}_2  \frac{\vec{\sigma}_1 \cdot \vec{q}\, \vec{\sigma}_2 \cdot \vec{q}}{\vec{q}\,^2 + m_\pi^2}  \, , \label{eqn:OPE}
\end{equation}
where the pion mass $m_\pi$ = 139.57 MeV, pion decay constant $f_\pi$ = 92.2 MeV and the axial coupling $g_A$ = 1.289. This potential is singular and attractive in $^3P_0$. When treated nonperturbatively, a counterterm $V_\text{ct}^{(0)}$, playing the role of short-range force, is needed at LO to renormalize OPE~\cite{Nogga:2005hy}:
\begin{equation}
  V_\text{LO}=V_{1\pi} + V_\text{ct}^{(0)} \, .
\end{equation}
The partial-wave decomposition of OPE onto $^3P_0$ is numerically performed with the following integral~\cite{Epelbaum-1999dj}:
\begin{equation}
    \langle ^3P_0; p | V_{1\pi} | ^3P_0; k \rangle =   -\frac{m_N}{4\pi^2}  \frac{g_A^2}{4f_\pi^2} \int_{-1}^{1} {\rm d}z\,\left[ 2pk-z(p^2+k^2)\right] \frac{1}{\vec{q}\,^2 + m_\pi^2}  \, ,
\end{equation}
where $\vec{q}\,^2 \equiv p^2 + k^2 - 2pk z$.

Together with consideration for multiple-pion exchanges \cite{Long:2011qx,Valderrama:2009ei}, nonperturbative OPE  induces more $\cp{3}{0}$ LECs at given order than NDA:
\begin{equation}
  \begin{split}
  \text{LO} :& \qquad V_\text{LO} = V_{1\pi} + C_0pp'\, ,\\
  \text{NLO} :& \qquad V_\text{NLO} = 0\, ,\\
  \text{N$^2$LO} :& \qquad V_\text{N$^2$LO} = V_{2\pi}^{(0)} + C_1pp'+D_0(p^2+p'^2)pp'\, ,\\
  \text{N$^3$LO} :& \qquad V_\text{N$^3$LO} = V_{2\pi}^{(1)} + C_2pp'+D_1(p^2+p'^2)pp'\, ,
  \end{split}
\label{eqn:npp}
\end{equation}
where $V_{2\pi}^{(0,\, 1)}$ are leading and subleading two-pion exchanges (TPEs). Matrix elements of $V_\text{ct}$ in $\cp{3}{0}$ are defined as
\begin{equation}
  \langle p' | V_\text{ct} | p \rangle = Cpp' + D (p^2+p'^2)pp' + E p^2 {p'}^2 p p' + \cdots\, ,
\end{equation}
with $Cpp'$ (and similarly other counterterms) being split into formally different pieces at each order \cite{Long:2011qx}
\begin{equation}
  Cpp' = (C_0 + C_1 + C_2 + ...)pp' \, ,
\end{equation}
where $C_1$ and $C_2$ are corrections to $C_0$ \cite{Fleming:1999ee,Long:2007vp}.

We are interested here in scenarios within chiral EFT that supports perturbation theory in $\cp{3}{0}$. The first scenario is to treat OPE in perturbation theory in $\cp{3}{0}$, arguing that the centrifugal barrier suppresses OPE so much to the point where $V_{1\pi}$ must be demoted to NLO while LO vanishes~\cite{Wu:2018lai}:
\begin{equation}
  \begin{split}
  \text{LO} :& \qquad V_\text{LO} = 0 \, ,\\
  \text{NLO} :& \qquad V_\text{NLO} = V_{1\pi}\, ,\\
  \text{N$^2$LO} :& \qquad V_\text{N$^2$LO} = C_0pp'\, ,\\
  \text{N$^3$LO} :& \qquad V_\text{N$^3$LO} = V_{2\pi}^{(0)} + C_1pp'\, ,\\
  \text{N$^4$LO} :& \qquad V_\text{N$^4$LO} = V_{2\pi}^{(1)} + C_2pp'+D_0(p^2+p'^2)pp'\, .
  \end{split}
\label{wu_long_pc}
\end{equation}
In Ref.~\cite{Wu:2018lai}, the above scheme was found to describe the phase shifts in $^3P_0$ well only below $k \simeq 180$ MeV, which is not significantly better than pionless EFT.

\begin{figure}[htbp]
\centering
\includegraphics[width=0.6\textwidth]{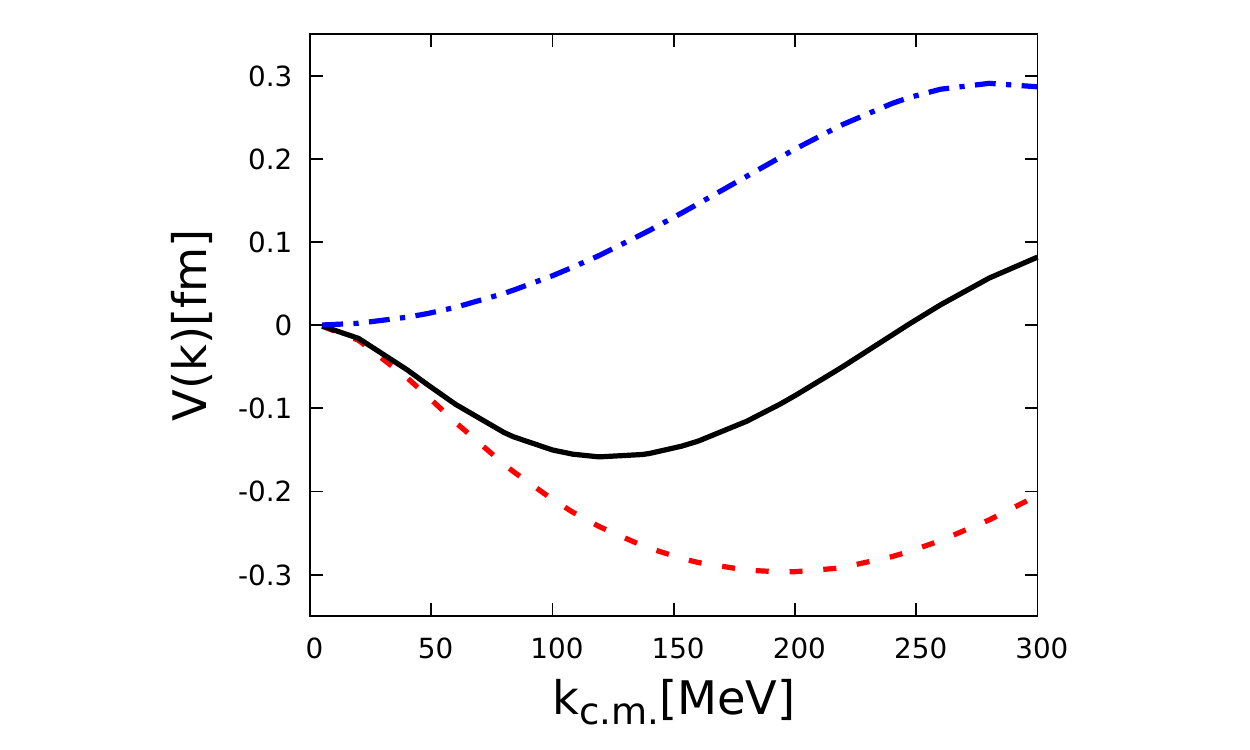}
\caption{On-shell matrix elements of $\cp{3}{0}$ potentials as functions of the center-of-mass momentum. The red dashed, blue dashed-dotted and black solid correspond respectively to $V_{1\pi}(k, k)$,  $C_0 k^2$ and $V_{1\pi}(k, k) + C_0 k^2$. $C_0$ is fitted in the nonperturbative OPE scheme~\eqref{eqn:npp} for $\Lambda = 400$ MeV. \label{fig:vkk400}}
\end{figure}

The second proposition, also the main message of the paper, is to examine the cancellation of the long and short-range forces in $\cp{3}{0}$, and explore the possibility that it supports perturbation theory of some kind. This is motivated by the observation that, at least for cutoff values $\Lambda$ not too high, the counterterm is repulsive. Shown in Fig.~\ref{fig:vkk400} are the on-shell matrix elements of potentials $V_{1\pi}(p,p')$, $ C_0pp'$ and $V_{1\pi}(p, p')$+$ C_0pp'$ as functions of $k$, where $p = p'$ for $\Lambda = 400$ MeV, and the value of $C_0$ is obtained with the nonperturbative OPE scheme~\eqref{eqn:npp}. The cancellation between $V_{1\pi}(p,p')$ and $C_0pp'$ is quite obvious: the maximum absolute value of $V_{1\pi}(p, p')$+$ C_0pp'$ is half of that of $V_{1\pi}(p,p')$. Although this is an encouraging sign, we need to go beyond the tree level and demonstrate the convergence of a perturbation theory in powers of $V_{1\pi}(p, p')$+$ C_0pp'$. To that end, we need a specific power counting scheme to account for the accidental smallness of the $\cp{3}{0}$ force before we can carry out actual calculations.

Unlike perturbative OPE scheme \eqref{wu_long_pc}, OPE is not considered to be suppressed by the centrifugal barrier for power-counting purpose, nor is the $\cp{3}{0}$ counterterm. In other words, $V_{1\pi}(p,p')$ and $ C_0pp'$ are still formally considered LO when separated, only the sum of the terms is considered to be NLO. The consequence of this thinking is that TPEs are not to be counted smaller than in NDA. As for counterterms with higher powers of momenta, we will let renormalization of multiple iteration of $V_{1\pi} + C_0pp'$ decide. That is, we add whatever counterterms necessary to remove divergence in the Born-expansion series of $V_{1\pi} + C_0pp'$. The divergences of the terms in the expansion can be straightforwardly estimated by powers of large loop momenta appearing in the integration. Up to \NNNLO, the aforementioned guideline leads to the following power counting:
\begin{equation}
  \begin{split}
  \text{LO} :& \qquad V_\text{LO} = 0\, ,\\
  \text{NLO} :& \qquad V_\text{NLO} = V_{1\pi} + C_0pp'\, ,\\
  \text{\NNLO} :& \qquad V_\text{\NNLO} = V_{2\pi}^{(0)} + C_1pp'+D_0(p^2+p'^2)pp'\, ,\\
  \text{\NNNLO} :& \qquad V_\text{\NNNLO} = V_{2\pi}^{(1)} + C_2pp'+D_1(p^2+p'^2)pp'+E_0p^2p'^2pp'\, .
  \end{split}
\label{eqn:perp}
\end{equation}
Here, renormalizing multiple iterations of $V_{1\pi} + C_0pp'$, in fact, brings about even more $\cp{3}{0}$ counterterms than the nonperturbative OPE ~\eqref{eqn:npp}, and scaling of contact operators do not agree with NDA. This counting is somewhat a middle ground between the perturbative OPE ~\eqref{wu_long_pc} and the nonperturbative OPE schemes.

\section{Results\label{sec:results}}

We show and discuss in this section the numerical results according to power counting ~\eqref{eqn:perp}. The expressions of TPEs are taken from Ref.~\cite{Epelbaum-1999dj}. The couplings of $\nu = 1$ $\pi \pi NN$ seagull vertexes called $c_i$'s determine the size of subleading TPEs $V_{2\pi}^{(1)}$, where $\nu$ is the chiral index ~\cite{Weinberg90, Weinberg91}. The value of $c_i$ are: $c_1=-0.74$, $c_3=-3.61$, $c_4=2.44$, all in GeV$^{-1}$, taken from Ref.~\cite{Siemens17}. They were extracted from an analysis based on the Roy-Steiner equation of $\pi N$ scattering. We use the NLO HB-NN $Q^2$ values of $c_i$ from Table I in Ref.~\cite{Siemens17}.

\begin{figure}[htbp]
\centering
\includegraphics[width=0.4\textwidth]{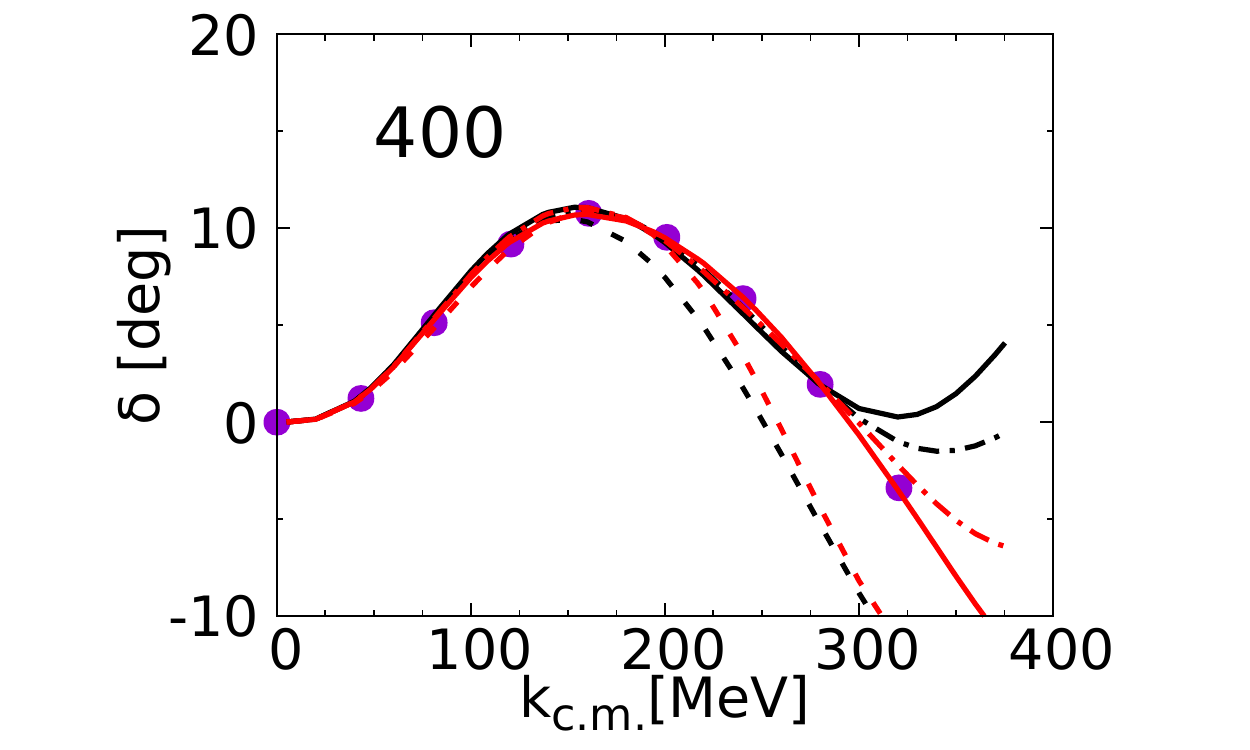}
\includegraphics[width=0.4\textwidth]{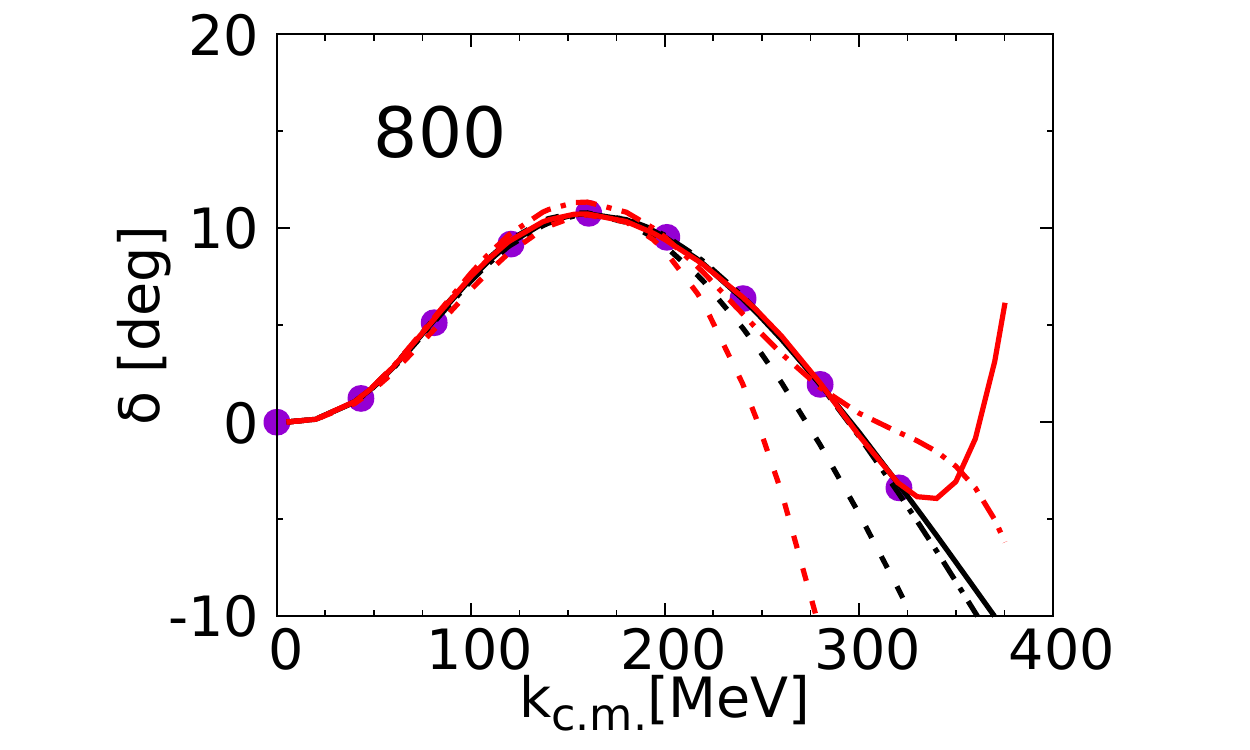}
\includegraphics[width=0.4\textwidth]{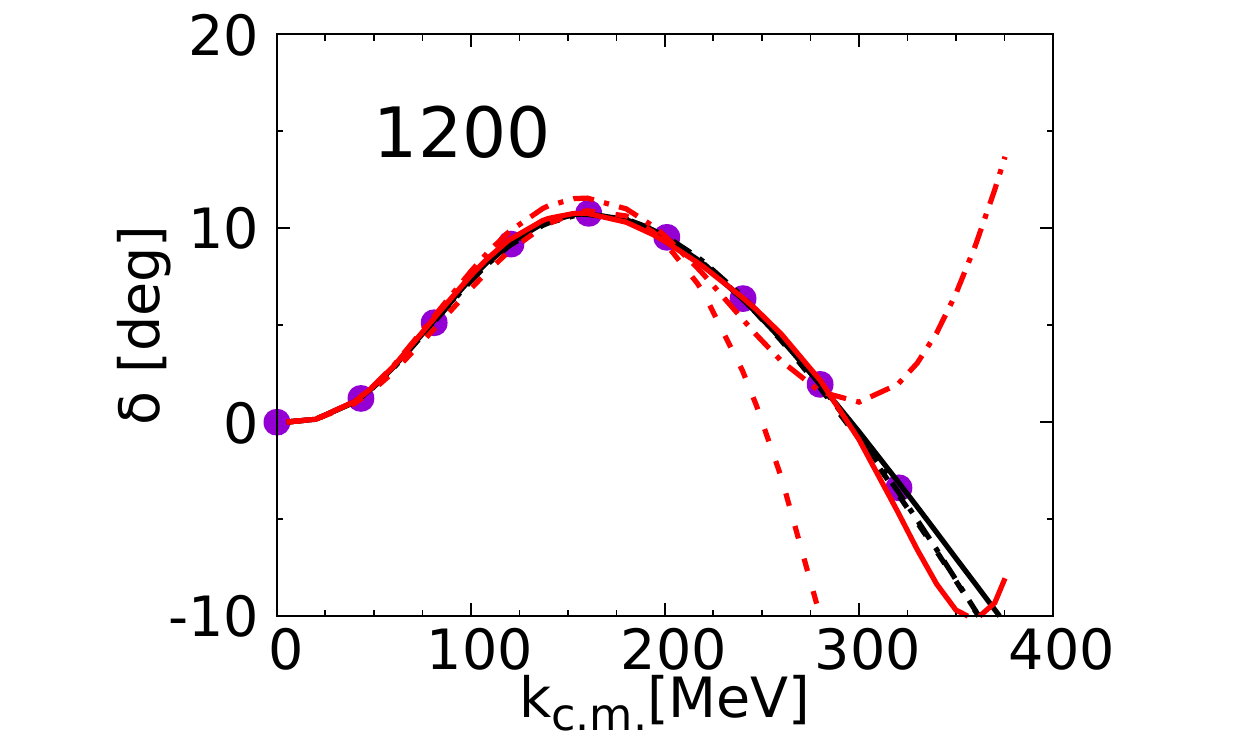}
\includegraphics[width=0.4 \textwidth]{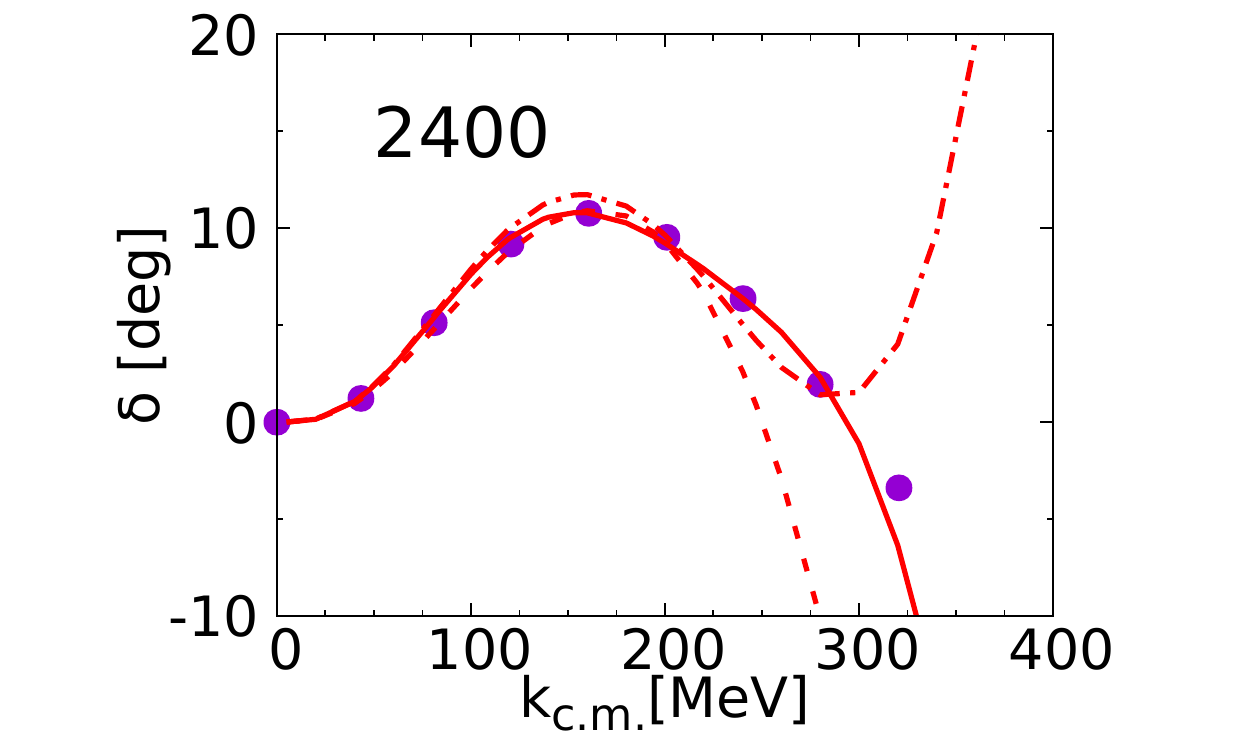}

\caption{The solid circles are the empirical phase shifts from the SAID program ~\cite{Arndt-2007qn, SAID}.The black and red lines correspond to the nonperturbative and perturbative cases, respectively. The dashed, dash-dotted and solid correspond respectively to LO, \NNLO~ and \NNNLO~ for the nonperturbative case, NLO, \NNLO~ and \NNNLO~ for the perturbative case. The values of cutoff $\Lambda$(MeV) are marked at top left of each figure.}
  \label{fig:pall}
\end{figure}

We fit the expanded amplitudes to the empirical phase shifts provided by the SAID program at the George Washington University (GWU)~\cite{Arndt-2007qn, SAID}. The results are shown in Fig.~\ref{fig:pall}. For NLO, phase shifts near $k = m_{\pi}$ are taken as inputs in the fitting. For \NNLO~and \NNNLO, phase shifts from threshold up to $k \simeq $ 300 MeV are used. For comparison, the nonperturbative scheme \eqref{eqn:npp} is also applied and shown, except for $\Lambda = 2400$ MeV where the algorithm we use fails to carry out higher-order calculations for the nonperturbative OPE scheme~\eqref{eqn:npp}. The values of the fitted counterterms are tabulated in Tables~\ref{tab:paranp} and ~\ref{tab:parahper}, where they are multiplied by powers of $\Lambda$ to be dimensionless.

\setlength{\tabcolsep}{+8pt}
\begin{table}
  \caption{The values taken by the counterterms in the nonperturbative OPE scheme~\eqref{eqn:npp}.}
  \label{tab:paranp}
  \centering
  \begin{tabular}{lcccccc}\\
  \hline
  \hline
  $\Lambda$ (MeV) \qquad & $C_0{\Lambda}^3$ \qquad & $C_1{\Lambda}^3$ \qquad & $D_0{\Lambda}^5$ \qquad & $C_2{\Lambda}^3$ \qquad & $D_1{\Lambda}^5$ \\
  \hline
  400.0 & $1.95$ & $1.35$ & $-0.258$ & $-0.680$ & $-1.47$  \\
  800.0 & $-22.2$ & $233$ & $-54.1$ & $201$ & $-61.1$ \\
  1200.0 & $-2.73$ & $58.2$ & $12.1$ & $-93.0$ & $-99.8$ \\
  \hline
  \hline
  \end{tabular}
\end{table}
\setlength{\tabcolsep}{-8pt}

\setlength{\tabcolsep}{+8pt}
\begin{table}
  \caption{The values taken by the counterterms in the perturbative scheme~\eqref{eqn:perp}.}
  \label{tab:parahper}
  \centering
  \begin{tabular}{lccccccc}\\
  \hline
  \hline
  $\Lambda$ (MeV) \qquad & $C_0{\Lambda}^3$ \qquad & $C_1{\Lambda}^3$ \qquad & $D_0{\Lambda}^5$ \qquad & $C_2{\Lambda}^3$ \qquad & $D_1{\Lambda}^5$ \qquad & $E_0{\Lambda}^7$ \\
  \hline
  400.0 & $1.79$ & $1.45$ & $-0.0330$ & $-0.242$ & $-2.87$ & $3.55$  \\
  800.0 & $15.4$ & $27.1$ & $32.9$ & $69.1$ & $-12.7$ & $516$  \\
  1200.0 & $51.0$ & $335$ & $536$ & $6.01\times10^3$ & $6.22\times10^3$ & $2.57\times10^4$ \\
  2400.0 & $408$ & $2.70\times10^4$ & $4.80\times10^4$ & $5.35\times10^6$ & $6.48\times10^6$ & $2.29\times10^7$ \\
  \hline
  \hline
  \end{tabular}
\end{table}
\setlength{\tabcolsep}{-8pt}

An important feature independent of the cutoff value is that the mixed perturbative scheme ~\eqref{eqn:perp} converges near $k \simeq 280$ MeV, whereas the pure OPE perturbative scheme ~\eqref{eqn:OPE} can describe the $\cp{3}{0}$ phase shifts only up to $k \simeq 180$ MeV~\cite{Wu:2018lai}. The improved convergence confirms our speculation: the cancellation between OPE and the $\cp{3}{0}$ survives in quantum fluctuations manifested by higher-order terms in the Born expansion~\eqref{eqn:amplitude}. Therefore, the perturbation theory based on
\begin{equation}
  V_\text{NLO} = V_{1\pi} + C_0pp'
\end{equation}
furnishes a more convergent EFT expansion than that based on
\begin{equation}
  V_\text{NLO} = V_{1\pi} \, .
\end{equation}

As successful as it is, the perturbation theory for $\cp{3}{0}$ still has a smaller convergence radius than the nonperturbative scheme~\eqref{eqn:npp} does, which has even fewer short-range parameters. Since OPE in $\cp{3}{0}$ will eventually become nonperturbative for sufficiently high momenta, the nonperturbative scenario is conceptually the underlying theory for the perturbative one. From this perspective, it would be less surprising that the nonperturbative theory "knows" more physics than the perturbative one, thus needs fewer parameters.

\begin{figure}[htbp]

\centering
\includegraphics[width=0.4\textwidth]{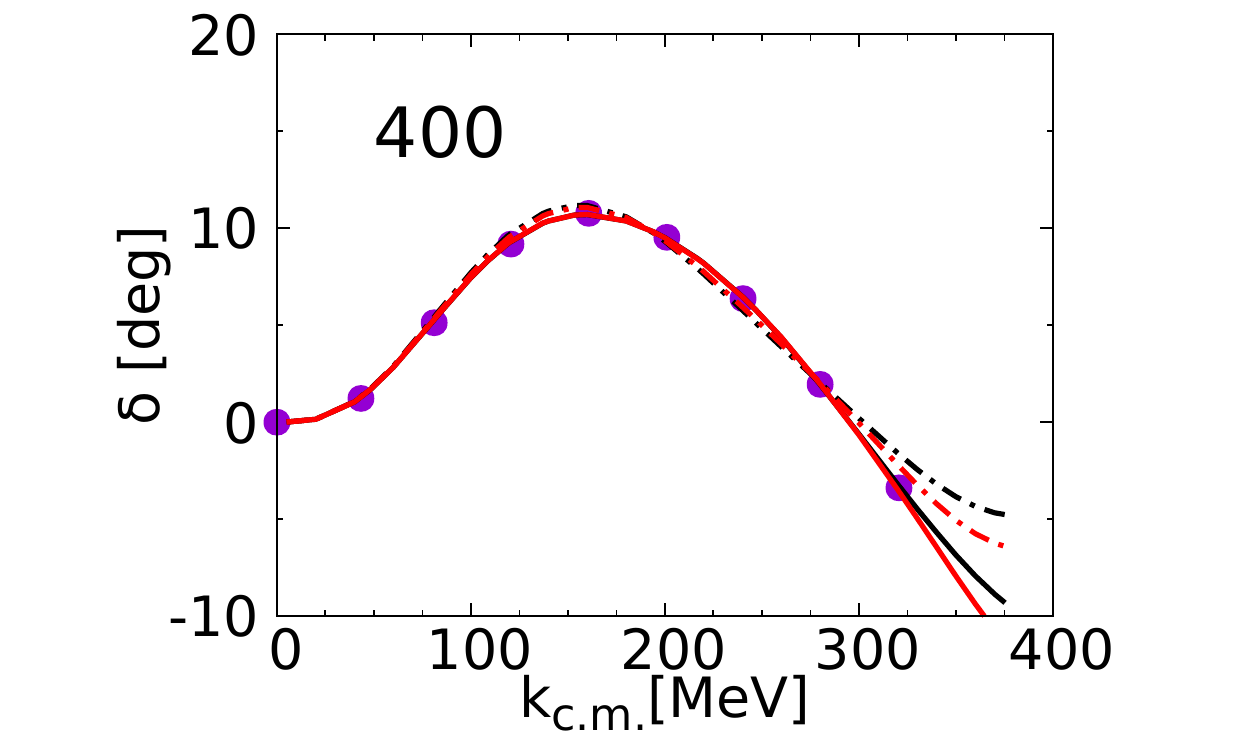}
\includegraphics[width=0.4\textwidth]{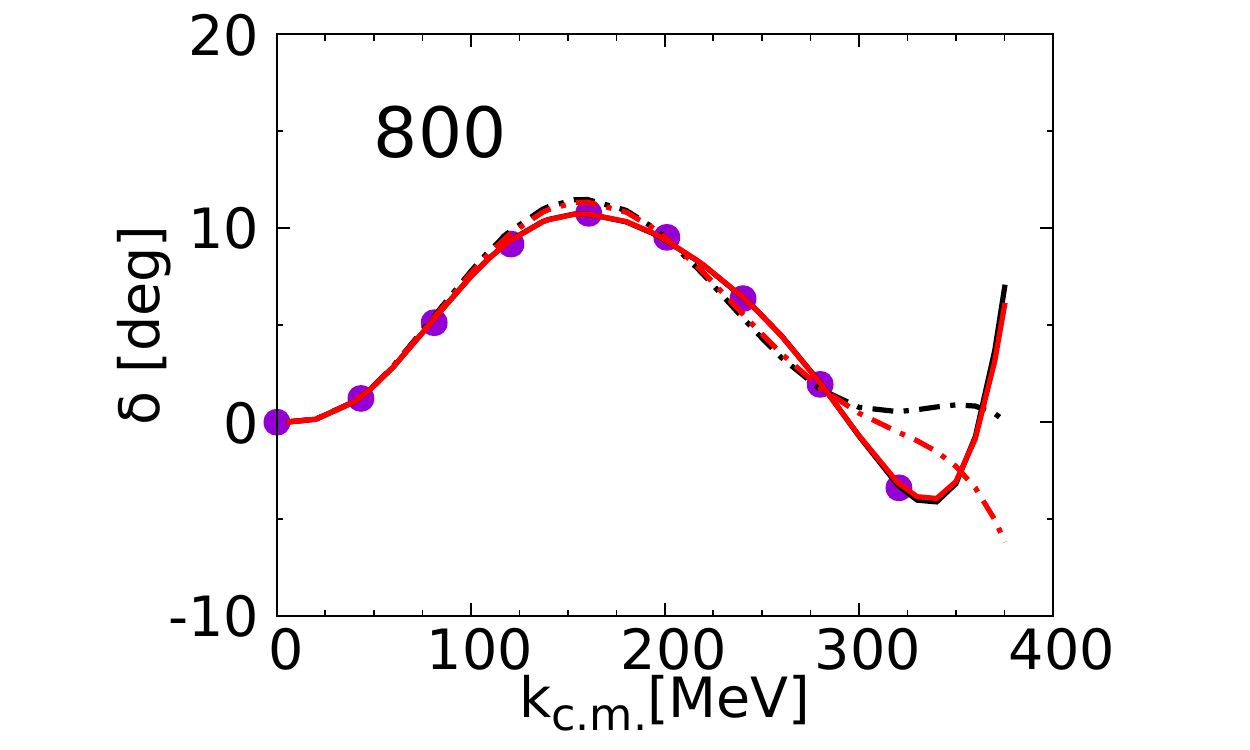}
\includegraphics[width=0.4\textwidth]{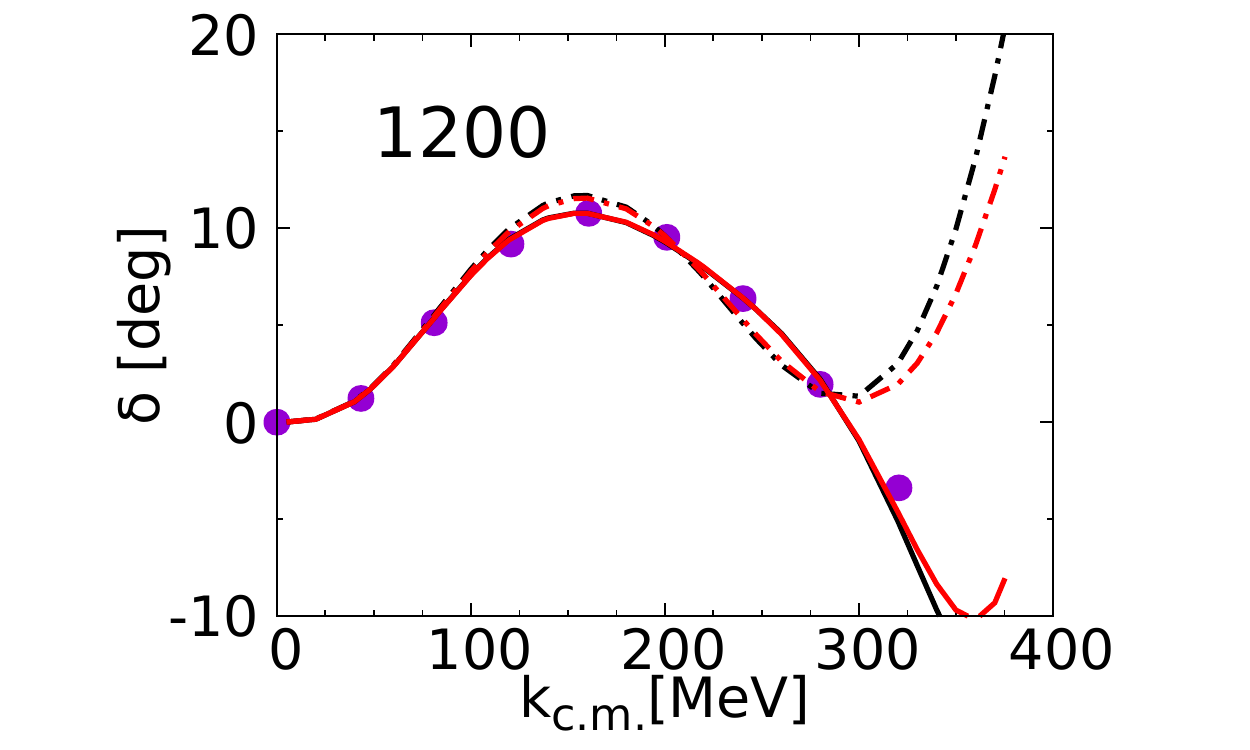}
\includegraphics[width=0.4\textwidth]{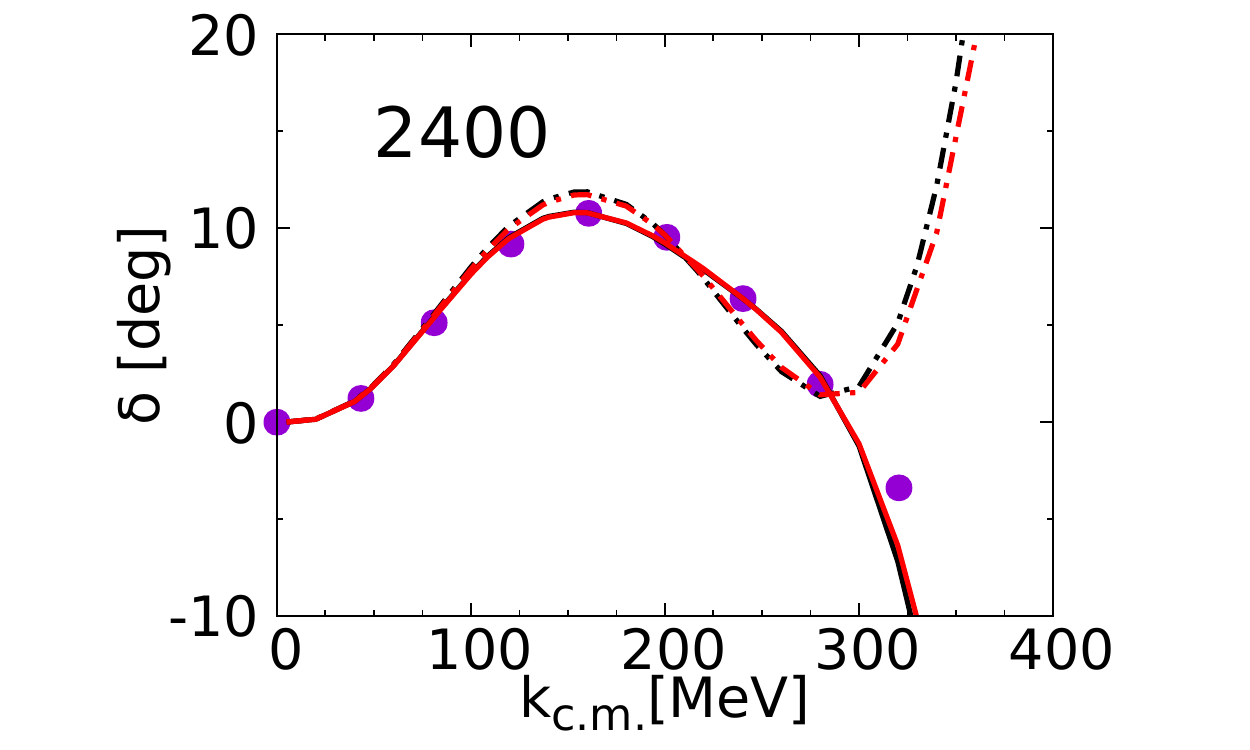}

\caption{The solid circles are the empirical phase shifts from the SAID program ~\cite{Arndt-2007qn, SAID}. The black and red lines correspond to the results with and without TPEs respectively. The dashed and solid correspond respectively to \NNLO~ and \NNNLO. The values of cutoff $\Lambda$(MeV) was marked at top left of each figure.}
  \label{fig:wot}
\end{figure}

One may ask whether it is TPEs or iterations of $V_{1\pi} + C_0pp'$ that drive the expansion scheme \eqref{eqn:perp} to break down near $k \simeq 280$ MeV. To answer this question, we ``turn off'' TPEs and investigate what happens:
\begin{equation}
  \begin{split}
  \text{LO} :& \qquad V_\text{LO} = 0\, ,\\
  \text{NLO} :& \qquad V_\text{NLO} = V_{1\pi} + C_0pp'\, ,\\
  \text{\NNLO} :& \qquad V_\text{\NNLO} = C_1pp'+D_0(p^2+p'^2)pp'\, ,\\
  \text{\NNNLO} :& \qquad V_\text{\NNNLO} = C_2pp'+D_1(p^2+p'^2)pp'+E_0p^2p'^2pp'\, .
  \end{split}
\label{wotp}
\end{equation}
The results are shown in Fig.~\ref{fig:wot}. The fact that the phase shifts with TPEs and without TPEs sit almost on top of each other tells us that the Born expansion of $V_{1\pi} + C_0pp'$, although softer than $V_{1\pi}$ alone, still becomes too strong for momenta where TPEs are expected to make impacts.

\section{Summary and Discussion\label{sec:sum}}

Given the relatively small values of the $\cp{3}{0}$ phase shifts, we were motivated to study perturbativeness of chiral potentials in $\cp{3}{0}$ by examining the cancellation between OPE and the short-range forces represented by the $\cp{3}{0}$ counterterms within the framework of chiral EFT. The numerical calculation showed that the perturbative scheme \eqref{eqn:perp} proposed in the paper can describe the empirical phase shifts up to $k \simeq 280$ MeV, where $k$ is the center-of-mass momentum.

As shown in Refs.~\cite{Wu:2018lai, Kaplan:2019znu}, OPE alone does not support a perturbation theory that would be as successful as the proposed perturbative scheme \eqref{eqn:perp}. So our exploration explains satisfactorily from the viewpoint of chiral EFT why $\cp{3}{0}$ phase shifts appear so small while OPE by itself is quite strong in $\cp{3}{0}$.


Since there is a shallow bound state in $\csd$, so the nuclear force must not be perturbative in $\csd$. But what prevents the same rationale employed in the paper to argue for perturbative $\cp{3}{0}$ being applied to $\csd$? After all, OPE is singularly attractive too in $\csd$ and the $\cs{3}{1}$ counterterm~\cite{Nogga:2005hy}. There is a fundamental difference between partial-wave projections of OPE in $\cp{3}{0}$ and $\csd$ that is often overlooked. The singular attraction of OPE acts in the orbital mixing, $\cs{3}{1} \to \cd{3}{1}$, but the counterterm is part of $\cs{3}{1} \to \cs{3}{1}$; therefore, they can not cancel each other at the tree level.

\acknowledgments

The work was supported in part by the National Natural Science Foundation of China (NSFC) under Grant Nos. 11775148 and 11735003.

\end{document}